# Scanning tunneling microscope characterizations of a circular graphene resonator realized with p-p junctions

Ya-Ning Ren[1, §], Jiao-Jiao Zhou[2, §], Zhong-Qiu Fu, Hai-Xuan Cheng[1], Si-Yu Li[1], Zi-Han Guo[1], Yi-Wen Liu[1], Chao Yan[1], Qi-Qi Guo[1], Jia-Bin Qiao[1], Yu Zhang[1], Sheng Han[1], Hua Jiang[2], and Lin He[1,3,*]

[1] Center for Advanced Quantum Studies, Department of Physics, Beijing Normal University, Beijing, 100875, People's Republic of China

[2] College of Physics, Optoelectronics and Energy and Institute for Advanced Study, Soochow University, Suzhou, 215006, People's Republic of China

[3] State Key Laboratory of Functional Materials for Informatics, Shanghai Institute of Mocrosystem and Information Technology, Chinese Academy of Sciences, 865 Changning Road, Shanghai 200050 , People's Republic of China

[§]These authors contributed equally to this work.

[*]Correspondence and requests for materials should be addressed to L.H. (e-mail: helin@bnu.edu.cn).

**Using low-temperature high-magnetic-field scanning tunneling microscopy and spectroscopy (STM/STS), we systematically study a graphene quantum dot (GQD) defined by a circular graphene *p-p* junction. Inside the GQD, we observe a series of quasi-bound states arising from whispering-gallery-mode (WGM) confinement of the circular junction and directly visualize these quasi-bound states down to atomic dimensions. By applying a strong magnetic field, a large jump in energy of the quasi-bound states, which is about one-half the energy spacing between the quasi-bound states, is observed. Such a behavior results from turning on a $\pi$ Berry phase of massless Dirac fermions in graphene by a magnetic field. Moreover, our experiment demonstrates that a quasi-bound state splits into two peaks with an energy separation of about 26 meV when the Fermi level crosses the quasi-bound state, indicating that there are strong electron-electron interactions in the GQD.**

The so-called Klein tunneling in graphene, which is characterized by perfect transmission for normal incidence and nearly perfect reflection for large-angle oblique incidence of massless Dirac fermions at *p-n* junctions, is one of the most exotic phenomena in solid-state systems [1-6]. Very recently, the Klein tunneling in graphene, i.e., the unusual anisotropic transmission, has been explored extensively in experiment [7-16]. For example, it was demonstrated explicitly that the nearly perfect reflection for oblique incidence leads to whispering-gallery-mode (WGM) confinement of the massless Dirac fermions and, consequently, results in the formation of quasi-bound states in circular graphene *p-n* junctions [7,9,10,12]. Meanwhile, the perfect transmission and no reflection for normal incidence enable turning on and off a π Berry phase of the temporarily confined massless Dirac fermions by a magnetic field, which results in a sudden and large jump in energy of the quasi-bound states in the circular graphene p-n junctions [14,15]. Perhaps because the fact that states at positive and negative energies are intimately linked is the essential feature responsible for the Klein tunneling in quantum electrodynamics (QED), therefore, only *p-n* junctions that interconnect electron and hole states in graphene are studied in experiment up to now. In this paper, we demonstrate that the unusual anisotropic transmission of the massless Dirac fermions (the Klein-like tunneling) also holds true for graphene *p-p* junctions. We systematically study electronic properties of a graphene quantum dot (GQD) defined by a circular graphene *p-p* junction using low-temperature high-magnetic-field scanning tunneling microscopy (STM). Both the quasi-bound states arising from the WGM-type confinement and the large jump in energy of the quasi-bound states induced by turning on the π Berry phase are observed in the GQD. Our experiment also demonstrates that a quasi-bound state splits into two peaks when the Fermi level crosses the quasi-bound state. Such a result indicates that strong electron-electron interactions may play an important role in affecting electronic properties of the GQD.

In our experiment, the graphene monolayer is directly synthesized on a Cu foil via a traditional low-pressure chemical vapor deposition method [17,18], and then the graphene sheet is transferred onto another Cu foil (see supplemental materials Figure S1 and S2 for details [19]). Previous studies indicate that the graphene monolayer directly synthesized on the Cu foil is usually *n* doped [10-12,16]. Introducing monolayer Cu vacancy island or intercalating monolayer S island between the graphene sheet and the substrate will locally enlarge distance between the graphene and the Cu foil and, as a consequence, change the graphene sheet to be *p* doped. Therefore, only graphene *n-p* junctions are studied previously [10-12,16]. However, for the graphene sheet transferred onto the Cu foil, two different features arising from the transfer process are obtained. First, the transfer process locally enlarges distance between the graphene sheet and the Cu foil, which makes large-area graphene sheet *p* doped. Then, the roughness of interface introduced by the transfer process can locally affect the doping of graphene and, consequently, generate *p-p* junctions in the graphene. Second, the transfer process reduces the interactions between the graphene sheet and the Cu foil, which enables us to locally measure Landau quantization of the graphene sheet in high magnetic field (see Figure S3 of supplemental materials [19]). Then, we can directly obtain the Dirac point of graphene according to the measured zero Landau level in high magnetic fields.

Figure 1(a) shows a representative STM image of a graphene area with a GQD defined by a circular *p-p* junction, as roughly indicated by the dashed curve. The boundary of the GQD is determined by carrying out spatial-resolved STS measurements (see Figure S4 of supplemental materials [19]) and the radius of the GQD *R* is estimated as about 13 nm. Figure 1(b) shows four representative STS spectra recorded at different positions of the studied area. Outside the GQD, the spectrum exhibits a typical V-shaped curve of graphene monolayer and the Dirac point is estimated as about 40 meV according to the zero Landau level obtained in high-magnetic-field measurements (see Figure S3 of supplemental materials [19]). Inside the GQD, the Dirac point moves to about 280 meV and a series of resonance peaks appear below the Dirac point in the tunneling spectra, indicating that the massless Dirac fermions are temporarily confined into quasi-bound states in the GQD. The average energy spacing of the quasi-bound states ~ 48 meV agrees well with that estimated according to $\Delta E \approx \hbar v_F / R$, where $\hbar$ is the reduced Planck's constant and $v_F = 1.0 \times 10^6$ m/s is the Fermi velocity of graphene monolayer (see Figure S5 of supplemental materials [19] for energy spacing of the quasi-bound states). The intensities of the quasi-bound states measured in the GQD depend sensitively on the recorded positions, as shown in Fig. 1(b) and 1(c). For example, the lowest quasi-bound state with the angular momentum $m = 1/2$ exhibits a maximum in the center of the GQD and the higher-energy resonances with higher-angular momentum show maxima near the boundary of the GQD. All the above results, including the energy spacing and the spatial distribution of the quasi-bound states, are similar as that observed in the GQD defined by a circular *p-n* junction [7,9,10,12], indicating the quantum confinement of the massless Dirac fermions in the GQD via the WGM confinement. Further inspection of the WGM confinement of the massless Dirac fermions in the GQD is measured by STS maps, which reflect the spatial distribution of the confined massless Dirac fermions. Figure 1(d) shows a representative STS map recorded at the energy of a quasi-bound state with higher-angular momentum. Obviously, it displays ring structure with the maximum near the boundary of the GQD, which is a characteristic feature of the WGM confinement of the massless Dirac fermions.

To further explore the quantum confinement of the massless Dirac fermions in the GQD defined by the circular *p-p* junction, we carry out theoretical calculations of a GQD in a continuous graphene sheet based on the lattice Green's function (see supplemental materials [19] for details) [20]. The radius of the GQD is $R = 13$ nm, the Dirac points on and off the GQD are 40 meV and 280 meV respectively, as schematically shown in Fig. 2(a). Figure 2(b) shows the calculated local density of states (LDOS) of the quasi-bound states as a function of energy and radial distance. The lowest quasi-bound state with the angular momentum $m = 1/2$ mainly locates in the center of the GQD and the LDOS of the higher-energy quasi-bound states (with higher-angular momentum) show maxima progressively approaching the boundary of the GQD. The formation of the quasi-bound states in the GQD via the WGM confinement is explicitly demonstrated by showing the spatial distribution of LDOS for a higher-energy quasi-bound state ($m = 5/2$), as shown in Fig. 2(c). The ring-like structure, which reproduces the result obtained in our experiment, is a characteristic feature of the WGM confinement of the massless Dirac fermions in the GQD (here we should point out that all the higher-energy resonances with higher-angular momentum exhibit similar ring-like structure). In Fig. 2(d), we show the simulated LDOS at several different positions inside the GQD and the energy separation of the quasi-bound states is estimated as about 50 meV. In the calculation, both the irregularities of the boundary of the GQD and the roughness of the potential inside the GQD, as observed in our experiment, are not taken into account. However, the simulated results, including the energy spacing and the spatial distribution of the quasi-bound states, capture well the main features obtained in our experiment.

In the GQDs defined by a circular *p-n* junction, it was demonstrated explicitly that a magnetic field will lead to a sudden and large jump in energy of the higher-angular-momentum quasi-bound states when the magnetic field reaches a critical value $B_C$ [14-16,21,22]. Such a behavior arises from the fact that the Berry phase of the quasiparticles confined in the GQDs can be switched on and off by the magnetic field because of the relativistic nature of the quasiparticles. In our experiment, we demonstrate that similar behavior could also be observed in the GQD defined by the circular *p-p* junction. Figure 3(a) shows two representative tunneling spectra measured near the edge of the GQD in zero magnetic field and in the magnetic field of 8 T, respectively. A large increase in energy of the quasi-bound states, which is about one-half the energy spacing between the quasi-bound states, is clearly observed when we measure the spectra in the high magnetic field. This experimental result can be understood in terms of the Berry phase [14-16,21,22] and the schematic figures shown in Fig. 3(b-d) provide an intuitive visualization of this phenomenon. The applied magnetic field bends the trajectories of the quasiparticles and changes their orbits. Above the critical magnetic field, the Lorentz force can twist the orbit with angular momentum antiparallel to the field into a skipping orbit with loops, as shown in Fig. 3(c). Then, the closed momentum space trajectories enclose the Dirac point (Fig. 3(d)), which adds a Berry phase of $\pi$ to them and shifts the energy levels of the quasi-bound states accordingly. The value of the critical magnetic field depends sensitively on the profile of the junction and, in our experiment, it can be theoretically estimated as about 4 T (see supplemental materials [19] for details). Therefore, it is reasonable to observe the jump in energy of the quasi-bound states in the magnetic field of 8 T. Besides the shifts of the energy levels of the quasi-bound states, there are several weak peaks in the tunneling spectrum recorded in the high magnetic field (Fig. 3(a)). These weak peaks are attributed to the Landau levels of graphene monolayer outside the GQD generated in the presence of large perpendicular magnetic fields.

In our experiment, one of the quasi-bound states in the GQD is located around the Fermi level and an unexpected feature of this quasi-bound state is observed, as shown in Fig. 4(a). When the Fermi level crosses the quasi-bound state, it splits into two peaks flanking the Fermi energy with an energy separation of ~26 meV. Such a behavior has never been reported before in the GQDs and reminds us the quantum Hall isospin ferromagnetic states in graphene, where the Landau level will split when it is partially filled [23-27]. For the quantum Hall ferromagnetism, exchange interactions are plausible origin of the splitting and external magnetic fields are necessary to realize such a state. Very recently, it was demonstrated explicitly that a van Hove singularity of twisted graphene bilayer (TGB) will also split when it is partially filled [28-32] and such a splitting is closely related to strongly correlated states, such as the Mott insulating phase and superconductivity, in the magic-angle TGB [33-35]. It is interesting to note that the energy separation of the two peaks in our experiment is of the order of the on-site Coulomb repulsion in the GQD, which can be estimated as $e^2/(4\pi\varepsilon R)$ ~ 30 meV. Here $e$ is the electron charge, $\varepsilon$ ~ $3\varepsilon_0$ is the effective dielectric constant of graphene [36] and $R$ is the radius of the GQD. At present, we do not know the exact reason of the splitting. However, the emergence of the splitting when the quasi-bound state is partially filled indicates that electron-electron interactions may play a vital role in affecting electronic properties of the GQD, especially when the quasi-bound state is around the Fermi level (Fig. 4(b) and 4(c)).

In conclusion, the quasi-bound states arising from the WGM-type confinement and the large jump in energy of the quasi-bound states induced by turning on the π Berry phase are observed in the GQD defined by the graphene *p-p* junctions. Moreover, our experiment also indicates that strong electron-electron interactions may play an importance role in affecting electronic properties of the GQD. Further experiments should be carried out to explore how the electron-electron interactions affect the WGM-type confinement of the massless Dirac fermions in the GQD.

## Acknowledgements

This work was supported by the National Natural Science Foundation of China (Grant Nos. 11974050, 11674029). L.H. also acknowledges support from the National Program for Support of Top-notch Young Professionals, support from "the Fundamental Research Funds for the Central Universities", and support from "Chang Jiang Scholars Program".

# Figures

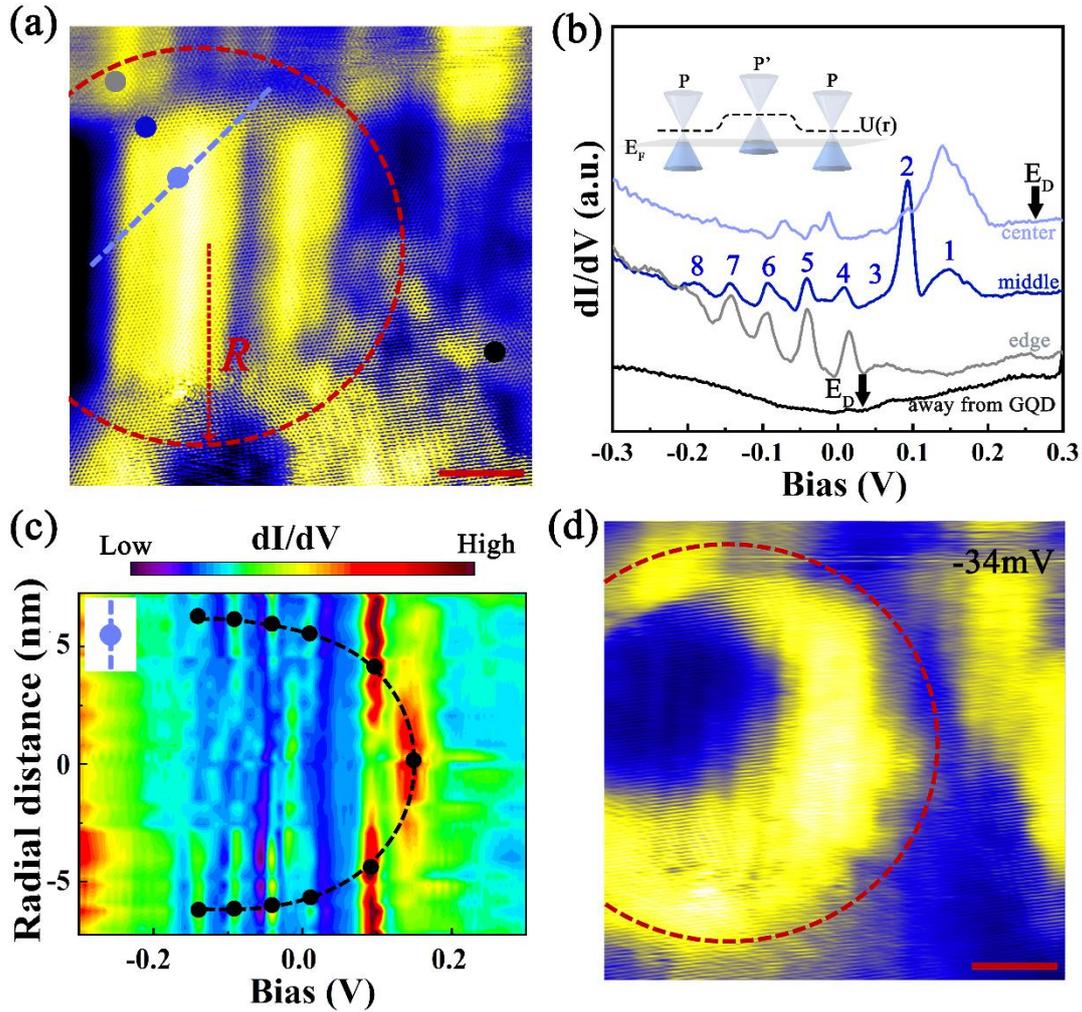

**FIG. 1. Imaging a circular graphene *p-p* junction.** (a) A 35 × 35 nm² STM topographic image ($V_{sample}$ = −34 mV, $I$ = 10 pA) showing a GQD on Cu substrate, as marked with the circular-shaped dashed curve. The scale bar is 5 nm. (b) Typical *dI/dV* spectra at different positions acquired on the dots with different colors in Fig. 1a. (a.u. = arbitrary units.). The arrows denote the positions of the Dirac points. The curves are offset on the Y axis for clarity. (c) Differential conductance map versus radial spatial position, corresponding to the light blue dashed line shown in (a). (d) *dI/dV* maps recorded around the GQDs with a fixed sample bias of -34 mV, which reveals the spatial distribution of the quasi-bound states. The scale bar is 5 nm.

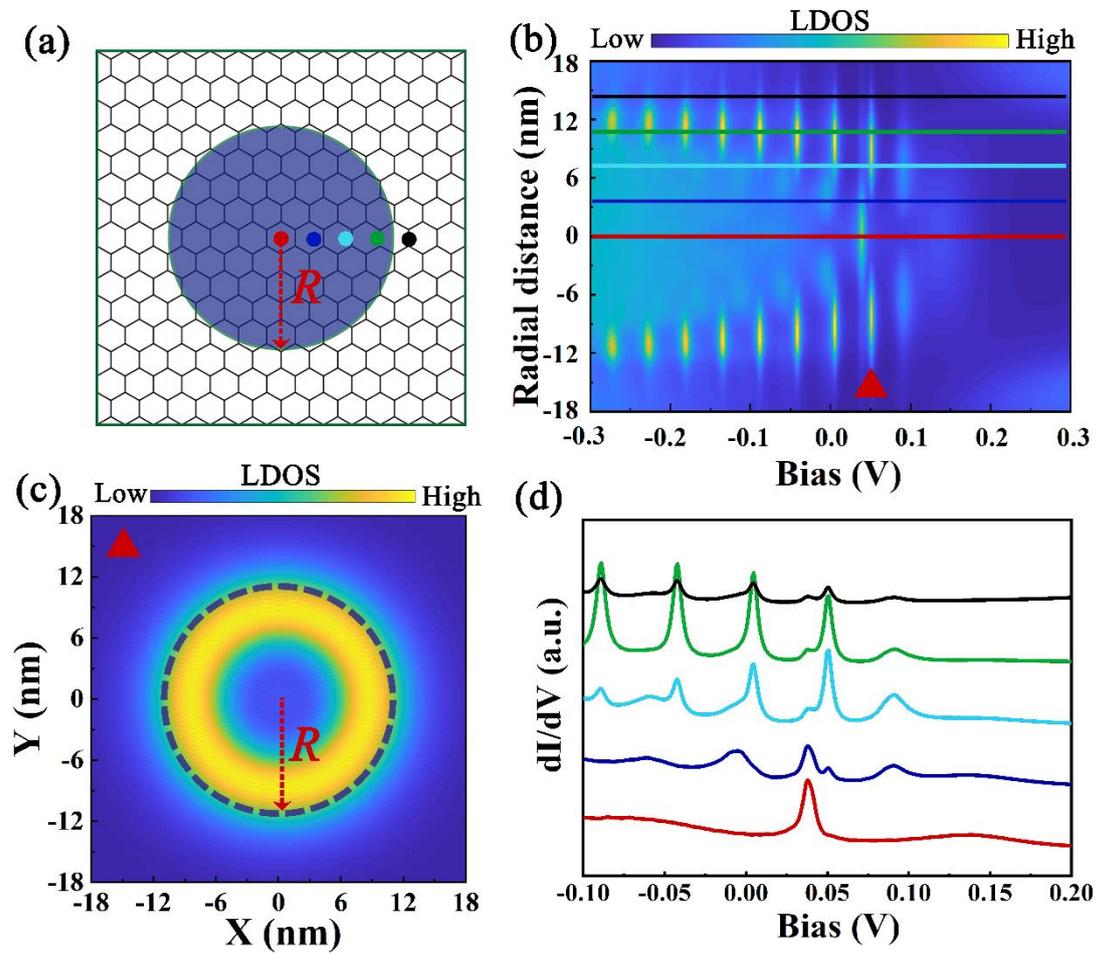

**FIG. 2. The theoretical simulation of quasibound states and the corresponding LDOS. (a)** Schematic structural model of a circular GQD. **(b)** Theoretically simulated LDOS as a function of bias and radial distance for the graphene resonator states. **(c)** Theoretical spatial distributions of the LDOS for the quasi-bound state as labelled in (b) (marked by the red triangle). **(d)** Calculated LDOS spectra corresponding to line cuts at radial distance indicated by solid lines in (b).

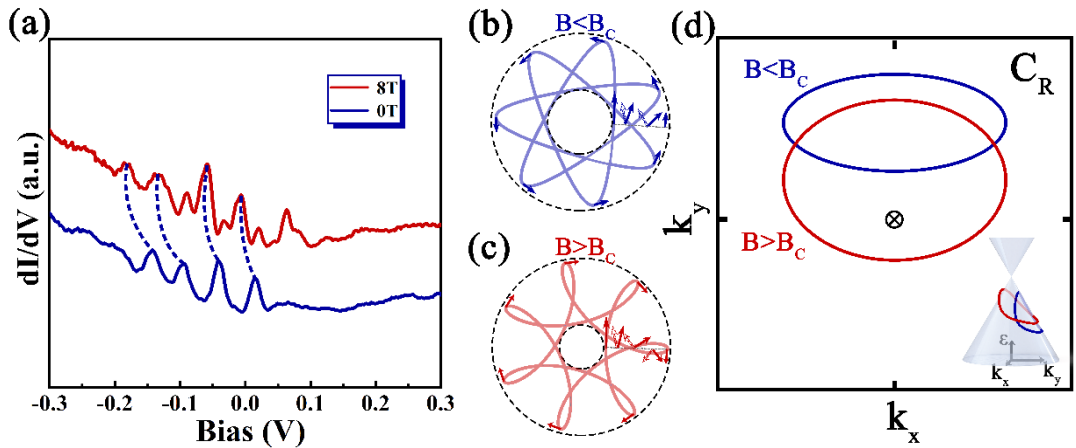

**FIG. 3. Effects of a large magnetic field on the quasi-bound states.** (a) STS spectra taken under an 8T field (red curve) and a 0T field (blue curve) at the edge of the GQD. The corresponding resonant peaks are marked by dashed curves. (b) and (c), Schematic diagrams of charge orbits in the GQD under different applied magnetic fields, corresponding to (b) $B < B_C$ and (c) $B > B_C$. (d) Schematic momentum-space contours for magnetic fields below (blue) and above (red) the critical magnetic field $B_C$. For above $B_C$ (red), there is a transition between winding of momentum, leading to a $\pi$ Berry phase. The inset is the charge trajectories in momentum space on the Dirac cone.

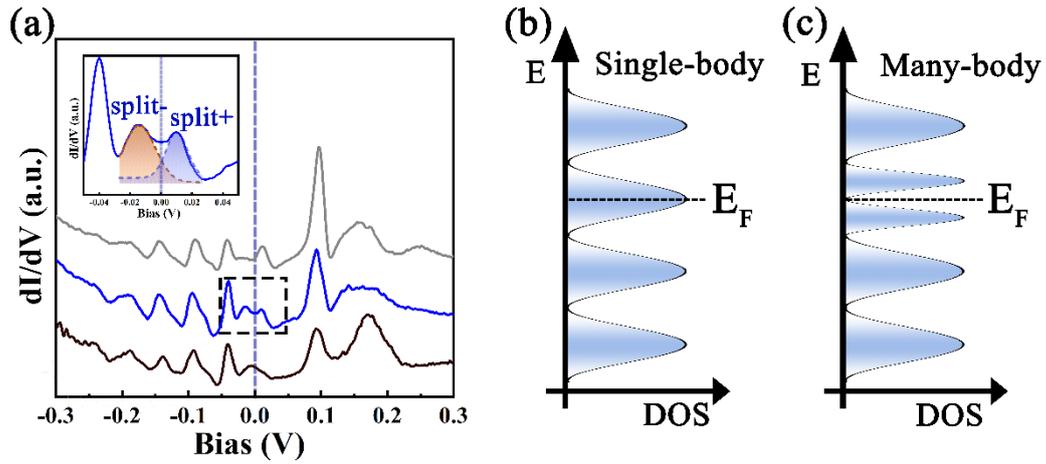

**FIG. 4. The splitting of quasi-bound states around the Fermi Level.** (a) The STS spectra of the quasi-bound states recorded at temperature 4K. One of the quasi-bound states splits into two peaks, one at -13 mV and the other at 13 mV. The inset shows the high-resolution spectrum of the splitting. (b) and (c) Schematic local density of states (LDOS) pictures. The single-particle quasi-bound state, as shown in (b), is split into upper and lower states due to the electron-electron interactions, as shown in (c).